\documentclass[aps,pre,twocolumn,showpacs,groupedaddress]{revtex4}

\usepackage{graphicx}
\usepackage{dcolumn}
\usepackage{bm}
\usepackage{amssymb}

\begin{document}

%\preprint{}

%Title of paper
\title{Swinging and Synchronized Rotations of Red Blood Cells in Simple Shear Flow}

\author{Hiroshi Noguchi}
\email[]{noguchi@issp.u-tokyo.ac.jp}
\affiliation{
Institute for Solid State Physics, University of Tokyo,
 Kashiwa, Chiba 277-8581, Japan}
\affiliation{
Institut f\"ur Festk\"orperforschung, Forschungszentrum J\"ulich, 
52425 J\"ulich, Germany}
%\homepage[]{Your web page}
%\altaffiliation{}

\date{\today}

\begin{abstract}
The dynamics of red blood cells (RBCs) in simple shear flow was studied using a 
theoretical approach based on three variables: a shape parameter, 
the inclination angle $\theta$, and phase angle $\phi$ of the membrane rotation.
At high shear rate and low viscosity contrast of internal fluid,
RBCs exhibit tank-treading motion, where $\phi$ rotates with swinging oscillation of shape and $\theta$. 
At low shear rate, tumbling motion occurs and $\theta$ rotates. 
In the middle region between these two phases,
it is found that synchronized rotation of $\phi$ and $\theta$ with integer ratios of the frequencies
occurs in addition to intermittent rotation.
These dynamics are robust to the modification of the potential of the RBC shape and membrane rotation.
Our results agree well with recent experiments.
\end{abstract}
%\pacs{87.16.D- Membranes, bilayers, and vesicles 
%      05.45.-a Nonlinear dynamics and chaos
%      82.40.Bj Oscillations, chaos, and bifurcations
%     }
%\keywords{}

\maketitle

\section{Introduction}

Soft deformable objects, such as liquid droplets, vesicles, cells, and  synthetic capsules
exhibit a complex behavior under flows. 
For example, in capillary flow,
fluid vesicles~\cite{vitk04}, red blood cells (RBCs)~\cite{fung97,fung04,skal69,gaeh80,nogu05b,mcwh09,pozr05}, 
and synthetic capsules~\cite{lefe08} 
deform to parachute shapes, and RBCs also deform to slipper shapes~\cite{skal69,gaeh80,nogu05b,mcwh09}.
Shape transitions of fluid vesicles occur in simple shear flow~\cite{nogu04,nogu05,nogu09}.
Membrane wrinkling appears
for fluid vesicles after inversion 
of an elongational flow~\cite{kant07,turi08} and for synthetic capsules in simple shear flow~\cite{walt01,fink06}.
Among these soft objects, RBC have received a great deal of attention,
since they are important  for both fundamental research and medical 
applications.  In microcirculation,
the deformation of RBCs reduces the flow resistance of microvessels.
In patients with diseases such as diabetes mellitus and sickle cell anemia, the RBCs have
a reduced deformability and often block the microvascular flow~\cite{fung97,tran84,nash83,tsuk01,higg07}.

In a simple shear flow with flow velocity ${\bf v}=\dot\gamma y {\bf e}_x$, 
fluid vesicles and RBCs show
a transition from a tank-treading (TT) mode with a constant inclination angle $\theta$ 
to a tumbling (TB) mode with increasing viscosity 
of the internal fluid $\eta_{\rm {in}}$~\cite{kell82,beau04,made06,kant06}
or membrane viscosity $\eta_{\rm {mb}}$~\cite{nogu04,nogu05}.
This transition 
is described well by the theory of Keller and Skalak 
(KS)~\cite{kell82}, which assumes a fixed ellipsoidal vesicle shape.
Experimentally, synthetic capsules and RBCs 
show the oscillation of their lengths and $\theta$, called swinging (SW) \cite{chan93,walt01,abka07},
during TT motion, and RBCs
also transit from TB to TT with increasing $\dot\gamma$  \cite{gold72,abka07}.
Recently, this dynamics was explained by the KS theory with the 
addition of an energy barrier for the TT rotation caused by the 
membrane shear elasticity~\cite{skot06,abka07}.
More recently, this transition was also obtained by simulations~\cite{kess08,sui08}.
However, the detailed dynamics has not yet been investigated.

For fluid vesicles in high shear flow,
shape transitions \cite{nogu04,nogu05,nogu09} occur,
and a swinging phase \cite{kant06,nogu07b,misb06,lebe07,lebe08}, 
where the shape and $\theta$ oscillate around $\theta \simeq 0$,
appears between the TT and TB phases.
This SW mode is also called trembling \cite{kant06,lebe07,lebe08} or vacillating-breathing \cite{misb06};
it is explained by
the KS theory extended to a deformable ellipsoidal vesicle \cite{nogu07b}
and the perturbation theory for a quasi-spherical vesicle \cite{misb06,lebe07,lebe08}.
Shape deformation plays an essential role in the SW of fluid vesicles.
The deformation is not necessary to explain
the SW of elastic capsules \cite{abka07,chan93,walt01,skot06,kess08,sui08,navo98,rama98}
but is required for quantitative analysis.
In this letter, we extend the theory in Ref.~\cite{skot06} to include the shape deformation of RBCs
and investigate the dynamics of deformable RBCs.

The internal fluid of RBCs behaves as a Newtonian fluid since
RBCs do not have a nucleus and other intracellular organelles.
The RBC membrane consists of a lipid bilayer with an 
attached spectrin network as cytoskeleton. 
The lipid bilayer is an area-incompressible fluid membrane. 
The shear elasticity of the composite membrane is induced by 
the spectrin network.
Under physiological conditions, 
an RBC has a constant volume $V = 94\mu{\rm m}^3$, surface area $S= 135\mu{\rm m}^2$,
 $\eta_{\rm {in}}=0.01$Pa$\cdot$s, $\eta_{\rm {mb}}\sim 10^{-7}-10^{-6}$Ns/m,
 membrane shear elasticity $\mu=6\times 10^{-6}$N/m, and 
bending rigidity $\kappa=2 \times 10^{-19}$J \cite{nogu09,fung04,moha94,tran84,dao06}.

The models and results are presented with dimensionless quantities (denoted by a superscript $*$).
The lengths and energies are normalized by $R_0=\sqrt{S/4\pi}$
and $\mu R_0^2$, respectively.
For RBCs, they are $R_0=3.3$ $\mu$m
and $\mu R_0^2=6.5\times 10^{-17}$J.
There are two intrinsic time units:
the shape relaxation time $\tau=\eta_0 R_0/\mu$ by the shear elasticity $\mu$,
and the time of shear flow $1/\dot\gamma$;
 the reduced shear rate is defined as $\dot\gamma^*=\dot\gamma \tau$.
The relative viscosities are
$\eta_{\rm {in}}^*=\eta_{\rm {in}}/\eta_0$
and $\eta_{\rm {mb}}^*=\eta_{\rm {mb}}/\eta_0R_0$, where 
$\eta_0$ is the viscosity of the outside fluid. 
In typical experimental conditions, the Reynolds number is low, Re$<1$; hence,
the effects of the inertia are neglected.

In Sec.~\ref{sec:fix}, we describe the extended KS theory~\cite{skot06} 
for an elastic capsule with a fixed ellipsoidal shape, and the phase behavior of the capsule.
In Sec.~\ref{sec:def}. we introduce the shape equation for deformable RBCs
and present the dynamics of deformed RBCs.
The dependence of the function shape of the RBC free-energy potential
is described in Sec.~\ref{sec:pot}.
Discussion and summary are given in Sec.~\ref{sec:dis} and Sec.~\ref{sec:sum},
respectively. 
The comparison with experimental results is presented in  Sec.~\ref{sec:dis}.

\section{Dynamics of elastic capsules with fixed shape}~\label{sec:fix}

\subsection{Models}

\subsubsection{Keller-Skalak Theory}

Keller and Skalak (KS) \cite{kell82}
 analytically derived the equation of the motion of vesicles or capsules
based on Jeffery's theory \cite{jeff22}.
In the KS theory,
the vesicles are assumed to have a fixed ellipsoidal shape, 
\begin{eqnarray}
\Big(\frac{x_1}{a_1}\Big)^2 +\Big(\frac{x_2}{a_2}\Big)^2 +\Big(\frac{x_3}{a_3}\Big)^2 =1, 
\end{eqnarray}
where $a_i$ are the semi-axes of the ellipsoid, and
the coordinate axes $x_i$ point along its principal directions. 
The $x_1$ and $x_2$ axes, with $a_1>a_2$, are on the vorticity ($xy$) plane,
and the $x_3$ axis is  in the vorticity ($z$) direction.
The maximum lengths in three directions are $L_1=2a_1$, $L_2=2a_2$, and $L_3=2a_3$.
The velocity field on the membrane is assumed to be 
\begin{eqnarray}
{\bf v}^{\rm {m}}=\omega {\bf u}^{\rm {m}}= 
               \omega \Big(-\frac{a_1}{a_2}x_2,\frac{a_2}{a_1}x_1,0\Big).
\label{eq:KS-vel}
\end{eqnarray}
The energy $W_{\rm {ex}}$ supplied from the external fluid 
has to be balanced with the energy dissipated in 
the vesicle, $W_{\rm {ex}}=D_{\rm {in}}+D_{\rm {mb}}$,
where $D_{\rm {in}}$ and $D_{\rm {mb}}$ are the energies dissipated
inside the vesicle and on the membrane, respectively. The motion of 
the vesicle is derived from this energy balance.
Then the motion of the inclination angle $\theta$ is given by
\begin{eqnarray}
\frac{d\theta}{dt} &=&  \frac{\dot\gamma}{2}\big\{-1+f_0 f_1 \cos(2\theta)\big\} - f_0 \omega \nonumber \\
 &=& \frac{\dot\gamma}{2}\{-1+B\cos(2\theta)\} 
\label{eq:thetb} \\
\label{eq:KS-B}
B &=& f_0\left\{f_1+ \frac{f_1^{-1}}
      {1+f_2(\eta_{\rm {in}}^* -1)
                  + f_2f_3 \eta_{\rm {mb}}^*}\right\}\\
\label{eq:KS-omega}
\omega &=& -\frac{\dot\gamma \cos(2\theta) }
      {2f_1\{1+f_2(\eta_{\rm {in}}^* -1) 
                  + f_2f_3 \eta_{\rm {mb}}^*\}}.
\end{eqnarray}
The membrane-viscosity term has been derived by Tran-Son-Tay {\it et al.} 
\cite{tran84}.
The factors appearing in Eqs.~(\ref{eq:thetb}-\ref{eq:KS-omega}) are 
given by
\begin{eqnarray*}
f_0 &=& 2/(a_1/a_2+a_2/a_1),\\
f_1 &=& 0.5(a_1/a_2-a_2/a_1),\\
f_2 &=& 0.5g(\alpha_1^2+\alpha_2^2),\\
f_3 &=& 0.5E_{\rm s}R_0/(f_1^2V),\\
g &=& \int_0^\infty (\alpha_1^2+s)^{-3/2}(\alpha_2^2+s)^{-3/2}
          (\alpha_3^2+s)^{-1/2}ds,\\
\alpha_i &=& a_i/(a_1a_2a_3)^{1/3},\\
E_{\rm s} &=& \oint \tilde{e}_{ij}\tilde{e}_{ij}dS,\\
\tilde{e}_{ij} &=& e_{ij}-0.5\Theta P_{ij},\\
e_{ij} &=& 0.5P_{ik}P_{jl}(\partial u^{\rm {m}}_k/\partial x_l
          +\partial u^{\rm {m}}_l/\partial x_k),\\
\Theta &=& P_{ij} \partial u^{\rm {m}}_i/\partial x_j,\\
P_{ij} &=& \delta_{ij}-n_in_j,
\end{eqnarray*}
where $E_{\rm s}$ is an integral over the membrane surface, and
${\bf n}$ is the normal vector of the surface.

For $B>1$, a stable fixed point $\theta=0.5\arccos(1/B)$ exists,
and TT motion occurs, while 
for $B<1$, there is no fixed point, and 
the angle $\theta$ periodically rotates (TB).
As $\eta_{\rm {in}}^*$ or $\eta_{\rm {mb}}^*$ increases,
the transition from TT to TB motion occurs,
where $B$ decreases from $B>1$ to $B<1$.
The membrane viscosity 
$\eta_{\rm {mb}}$ and the internal viscosity $\eta_{\rm {in}}^*$
have a similar effect; hence,
 an effective internal viscosity can be defined as
$\eta_{\rm {eff}}^*=\eta_{\rm {in}}^* +  f_3 \eta_{\rm {mb}}^*$.
The factor $f_3$ in $\eta_{\rm {eff}}^*$ depends on 
the vesicle shape and can give different dynamics for
deformable vesicles, in particular for shape transformations between
prolate and oblate vesicles \cite{nogu04,nogu05}.

The KS theory quantitatively predicts the TT-TB transition with  increasing $\eta_{\rm {eff}}^*$.
However, it cannot explain the TB-TT transition with increasing $\dot\gamma$.
In the KS theory, 
vesicle motion does not depend on $\dot\gamma$ except that
the TT or TB rotation velocity increases linearly with $\dot\gamma$.

\begin{figure}
\includegraphics{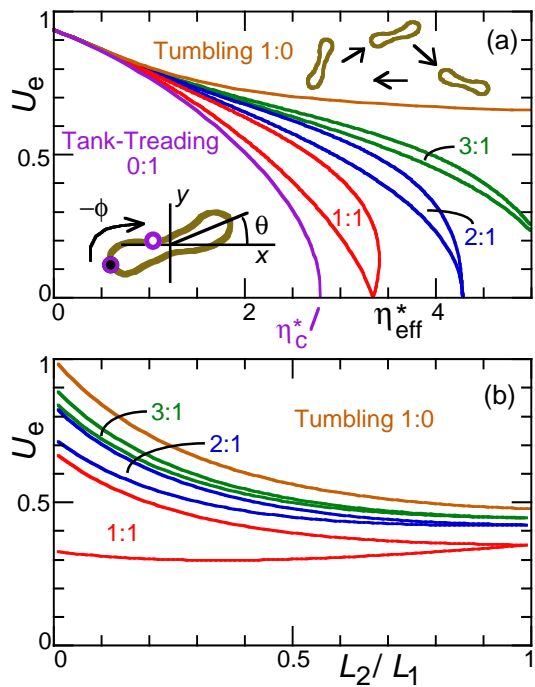}
\caption{ \label{fig:skot_phase}
(Color online)
Dynamic phase diagram of elastic capsules 
with fixed shapes calculated from Eqs.~(\ref{eq:qks}) and (\ref{eq:phiks}) with the potential $F_0=E_0\sin^2(\phi)$.
(a) Viscosity $\eta_{\rm {eff}}^*$ dependence for the RBC-like oblate shape with $L_2/L_1=0.25$ and $L_3/L_1=1$.
(b) Aspect ratio $L_2/L_1$ dependence at the TT-TB transition viscosity 
$\eta_{\rm {eff}}^*=\eta_{\rm {c}}^*$ of the KS theory ($U_{\rm e}=0$).
The boundary lines of the TB (brown), 
TT (violet), and synchronization regions 
[$f_{\rm {rot}}^{\theta}:f_{\rm {rot}}^{\phi}=1:1$ (red), $2:1$ (blue), and $3:1$ (green)] are shown.
}
\end{figure}

\subsubsection{KS Theory with an Energy Barrier}

Skotheim and Secomb extended the KS theory to take into account an energy barrier during TT membrane rotation~\cite{skot06}.
For RBCs and synthetic capsules with non-spherical rest shape, their membranes are locally deformed
during the TT rotation.
Fischer experimentally demonstrated that the RBC membrane rotates back to the original position 
when the shear flow is switched off~\cite{fisc04}.
To describe the energy barrier,  
a phase angle $\phi$ and free energy potential $F(\phi)$ are introduced; see inset of Fig.\ref{fig:skot_phase}.
The potential is periodic, $F(\phi+ n\pi)=F(\phi)$ and  $\phi=0$ at the rest shape.
Thus, the motions of the inclination angle
$\theta$ and phase angle $\phi$ are given by
\begin{eqnarray}
\label{eq:qks}
\frac{\ \ d \theta}{\dot\gamma dt} &=& \frac{1}{2}\big\{-1+f_0 f_1 \cos(2\theta)\big\} - \frac{f_0 d \phi}{\dot\gamma dt},\\
\label{eq:phiks}
\frac{\ \ d \phi}{\dot\gamma dt} &=& -\frac{(c_0/\dot\gamma^*V^*) \partial F^*/\partial \phi + \cos(2\theta) }
      {2f_1\{1+f_2(\eta_{\rm {in}}^* -1) 
                  + f_2f_3 \eta_{\rm {mb}}^*\}},
\end{eqnarray}
where $c_0=3f_2/8\pi f_1$.
The equations of the original KS theory are recovered
 in the absence of barriers of the free energy $F$,
i.e., $\partial F^*/\partial \phi=0$,
where $\omega=d\phi/dt$ is independent of $\phi$.

Skotheim and Secomb used a simple potential $F_0(\phi)=E_0 \sin^2(\phi)$
and a reduced energy $U_{\rm e}=f_2E_0/2f_1\eta\dot\gamma V = E_0^* c_0/\dot\gamma^*V^*$.
We employ the potential $F_0$ in this section
and describe the dependence on the potential shape in Sec.~\ref{sec:pot}.
Eqs.~(\ref{eq:qks}) and (\ref{eq:phiks}) are numerically integrated
using the fourth-order Runge-Kutta method.
An oblate capsule with  $L_2/L_1=0.25$ and  $L_3/L_1=1$
 is used as a model RBC.

\begin{figure}
\includegraphics{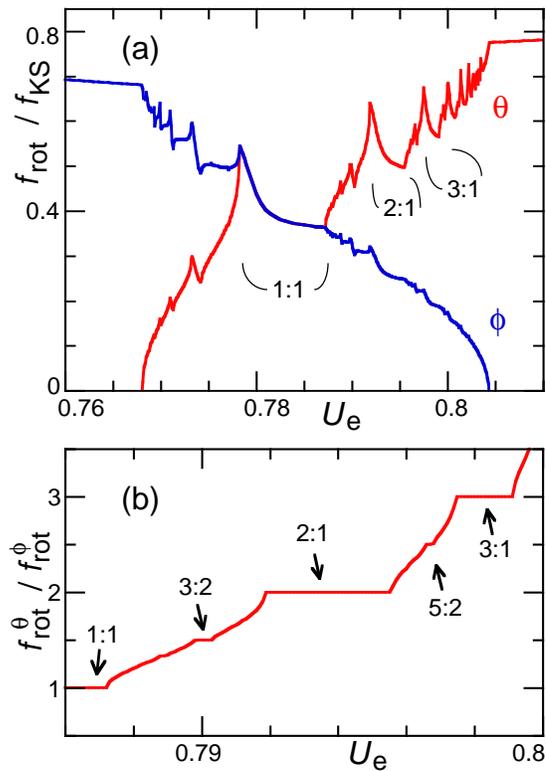}
\caption{ \label{fig:skot_frot}
(Color online)
Rotation frequency $f_{\rm {rot}}$ of 
the inclination angle $\theta$ and phase angle $\phi$
for the fixed oblate shape with $L_2/L_1=0.25$ 
at $\eta_{\rm {eff}}^*=1$.
Numbers represent $f_{\rm {rot}}^{\theta}:f_{\rm {rot}}^{\phi}$.
}
\end{figure}

\begin{figure}
\includegraphics{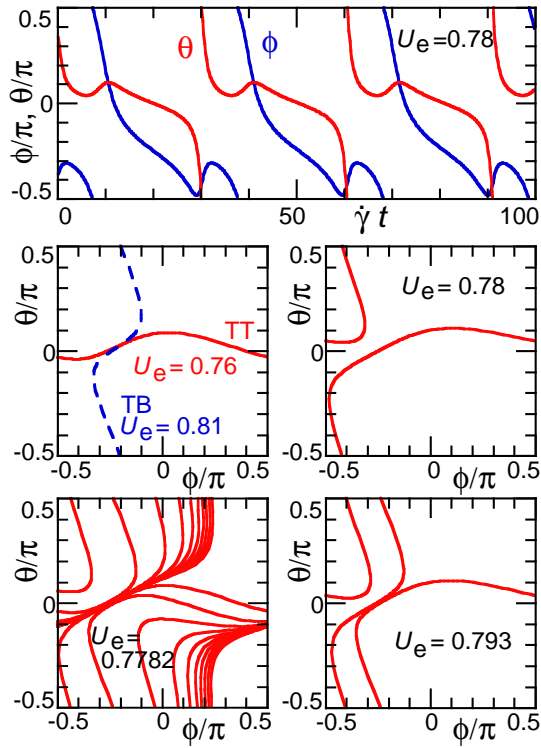}
\caption{ \label{fig:skot_det}
(Color online)
Dynamics of oblate capsules for the fixed shape.
Top panel: time development of angles $\theta$ and $\phi$ at $U_{\rm e}=0.78$.
Middle and bottom panels: trajectories on the phase space ($\theta$, $\phi$).
The capsules exhibit TT and TB rotations at  $U_{\rm e}=0.76$ and $0.81$, respectively.
Synchronized rotations with $f_{\rm {rot}}^{\theta}:f_{\rm {rot}}^{\phi}=1:1$
and $2:1$, and intermittent rotation are observed at $U_{\rm e}=0.78$, $0.793$, and $0.7782$, 
respectively.
The other parameters are the same as in Fig.~\ref{fig:skot_frot}.
}
\end{figure}

\begin{figure}
\includegraphics{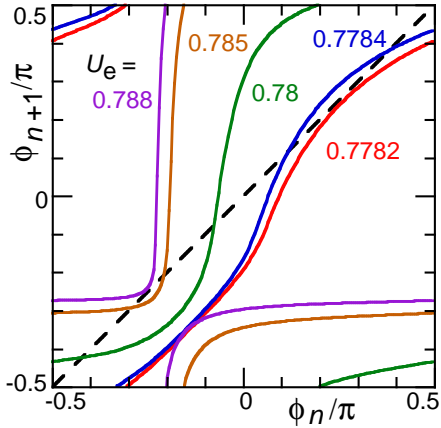}
\caption{ \label{fig:skot_rmap}
(Color online)
Return map of oblate capsules for the fixed shape.
The angle $\phi$ at $\theta=\pi/2- n \pi$ is plotted.
Synchronized rotation with $f_{\rm {rot}}^{\theta}:f_{\rm {rot}}^{\phi}=1:1$
occurs at $0.77823 <U_{\rm e}<0.78722$.
A dashed line represents $\phi_{n+1}=\phi_n$.
The other parameters are the same as in Fig.~\ref{fig:skot_frot}.
}
\end{figure}

\subsection{Results}

Figures \ref{fig:skot_phase}-\ref{fig:skot_rmap} show the dynamics of elastic capsules with fixed shapes.
At low shear rate $\dot\gamma$ (large $U_{\rm e}$), 
the capsules show TB motion,
since the free energy barrier locks the phase angle at $\phi\simeq 0$.
In TB, $\theta$ rotates but $\phi$ osculates; see the trajectory at $U_{\rm e}=0.81$ in Fig.~\ref{fig:skot_det}.
At higher $\dot\gamma$ (smaller $U_{\rm e}$) and low $\eta_{\rm {in}}$, 
TT motion occurs.
In TT,  $\theta$ oscillates (swing), and
$\phi$ rotates instead of $\theta$;
 see the trajectory at $U_{\rm e}=0.76$ in Fig.~\ref{fig:skot_det}.
The oscillation of $\phi$ or $\theta$ in TB or TT occurs with the rotation frequency $f_{\rm {rot}}^{\theta}$ or 
$f_{\rm {rot}}^{\phi}$, respectively.
Here, an angle change of $\pi$ is counted as one rotation.
Skotheim and Secomb~\cite{skot06} reported an intermittent phase between the TT and TB phases,
where both rotations of  $\theta$ and $\phi$ occur.
The $\phi$ (TT) rotation is intermittently interrupted by the $\theta$ (TB) rotation
slightly above the maximum energy barrier $U_{\rm e}^{\rm {tt}}$ of the TT phase.
However, we found that the phases of synchronized rotation of $\theta$ and $\phi$ also exist in this middle range between the TB and TT phases.
An infinite number of synchronization phases with integer ratios of $f_{\rm {rot}}^{\theta}$ and $f_{\rm {rot}}^{\phi}$ exist; see Fig.~\ref{fig:skot_frot}. 
This type of synchronization is called the Devil's staircase \cite{berg84}.

The trajectories of the synchronized rotations with $f_{\rm {rot}}^{\theta}:f_{\rm {rot}}^{\phi}=1:1$ and $2:1$
are shown in Fig.~\ref{fig:skot_det}.
The former
has the widest $U_{\rm e}$ range.
The approach to synchronization is explained by the return map $\phi_{n+1}(\phi_n)$ 
at $\theta=\pi/2- n \pi$ in Fig.~\ref{fig:skot_rmap}.
The curve $\phi_{n+1}(\phi_n)$ shifts to the left with increasing $U_{\rm e}$
and has two (stable and unstable) crossing points 
with the line $\phi_{n+1}=\phi_n$ at $0.77823 <U_{\rm e}<0.78722$, where
the capsules approach the limit cycle with $\phi_{n+1}=\phi_n$.
At slightly below or above the synchronization region ($0<0.77823 -U_{\rm e} \ll 1$ or $0<U_{\rm e} -0.78722 \ll 1$), 
intermittent rotation appears, like near the TT and TB regions~\cite{skot06};
see the trajectory at $U_{\rm e}=0.7782$ in Fig.~\ref{fig:skot_det}.

The qualitative behavior of capsules does not depend on the aspect ratios $L_2/L_1$ and $L_3/L_1$.
However, the ranges of the synchronized rotations are narrower for more spherical capsules, as shown in Fig.~\ref{fig:skot_phase}(b). 
Thus, it would be difficult to observe synchronized rotations in quasi-spherical capsules.
Note that the dynamics is independent of $L_3/L_1$ at $\eta_{\rm {eff}}^*=\eta_{\rm c}^*$ ($B=1$).

A fluid vesicle has no membrane shear elasticity ($U_{\rm e}=0$), 
and shows no synchronization between $\theta$ and $\phi$ since the potential does not depend on $\phi$;
see Fig.~\ref{fig:skot_phase}(a). 
At $0<\eta_{\rm {eff}}^*-\eta_{\rm c}^* \ll 1$, intermittent $\theta$ rotation also appears for the fluid vesicle.
The tumbling frequency is given by $f_{\rm {rot}}^{\theta}=\dot\gamma\sqrt{1-B^2}/2\pi \propto \sqrt{\eta_{\rm {eff}}^*-\eta_{\rm c}^*}$ 
in the KS theory~\cite{kell82}.
A similar frequency dependence on $U_{\rm e}$ is obtained in the extended KS theory:
  $f_{\rm {rot}}^{\theta} \propto \sqrt{U_{\rm {e}}-U_{\rm e}^{\rm {tt}}}$
slightly above $U_{\rm e}^{\rm {tt}}$~\cite{skot06}.

\begin{figure}
\includegraphics{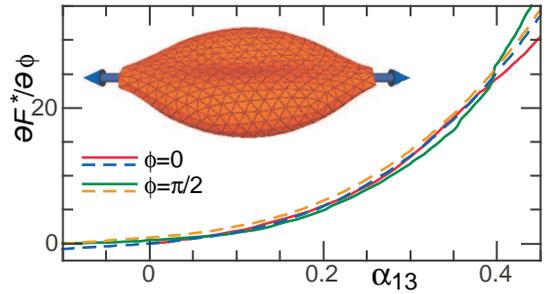}
\caption{ \label{fig:stretch}
(Color online)
Effective force  $\partial F^*(\alpha_{13},\phi)/\partial \phi$ at $\phi=0$ and $\pi/2$.
Solid and dashed lines represent the simulation data and fit functions, respectively.
In the inset,
a snapshot of the model RBC elongated by mechanical forces is shown.
}
\end{figure}

\section{Dynamics of deformable RBCs}~\label{sec:def}

\subsection{Shape equation}

Previously, we extended the KS theory to include the shape deformation of fluid vesicles, 
 on the basis of the perturbation theory \cite{misb06,lebe08,seif99} of quasi-spherical vesicles \cite{nogu07b}.
It showed very good agreement with experimental data \cite{kant06}.
Here, we have adapted it to RBC dynamics.
The shape parameter $\alpha_{12}=(L_1-L_2)/(L_1+L_2)$ does not increase monotonically for elongation,
because of RBC dimples.
Therefore, the shape parameter $\alpha_{13}=(L_1-L_3)/(L_1+L_3)$ is employed,
where $\partial \alpha_{13}/\partial \alpha_{12} = 2$ for an oblate ellipsoid ($\alpha_{13}=0$).
The equation of the shape evolution is given by
\begin{equation}
\label{eq:al}
\frac{d \alpha_{13}}{\dot\gamma dt} = \Big\{1-\big(\frac{\alpha_{13}}
                   {\alpha_{13}^{\rm {max}}}  \big)^2\Big\}
    \Big\{ -\frac{A_0}{\dot\gamma^*} 
       \frac{\partial F^*}{\partial \alpha_{13}} 
                + A_1\sin(2\theta)\Big\},
\end{equation}
where $A_0= 45/2\pi(32+23\eta_{\rm {in}}^*+16\eta_{\rm {mb}}^*)V^*$ and 
$A_1= 60/(32+23\eta_{\rm {in}}^*+16\eta_{\rm {mb}}^*)$.
Here, the terms of $\eta_{\rm {mb}}^*$ are added in $A_0$ and $A_1$
  based on the theory in Refs. \cite{lebe07,lebe08}. This revision
  improves the $\eta_{\rm {mb}}^*$ dependence of fluid
  vesicles in Ref. \cite{nogu07b}; the phase diagram for $\eta_{\rm {mb}}^*$ becomes similar to that for
  $\eta_{\rm {in}}^*$.

\begin{figure}
\includegraphics{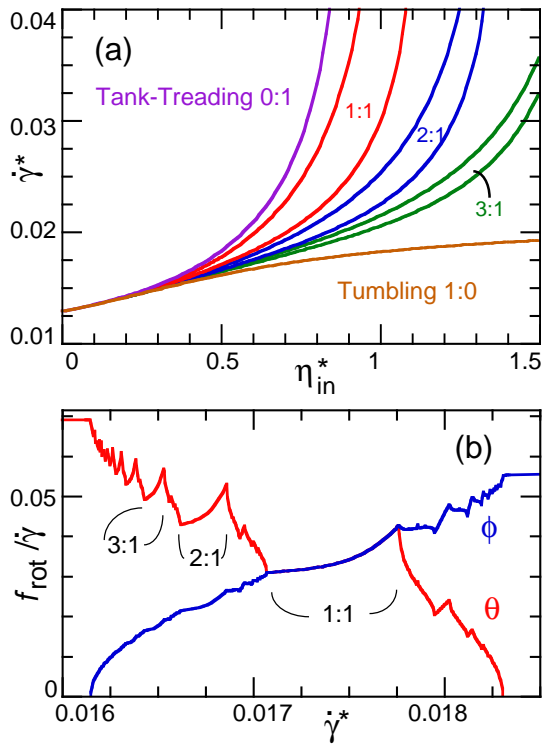}
\caption{ \label{fig:phase_std}
(Color online)
Dynamics of the deformable RBCs in the simple shear flow calculated from
Eqs.~(\ref{eq:qks}), (\ref{eq:phiks}), and (\ref{eq:al}).
(a) Dynamic phase diagram.
(b) Rotation frequency $f_{\rm {rot}}$ of 
$\theta$ and $\phi$
at $\eta_{\rm {in}}^*=0.5$ ($\eta_0= 0.02$Pa$\cdot$s ).
}
\end{figure}

\begin{figure}
\includegraphics{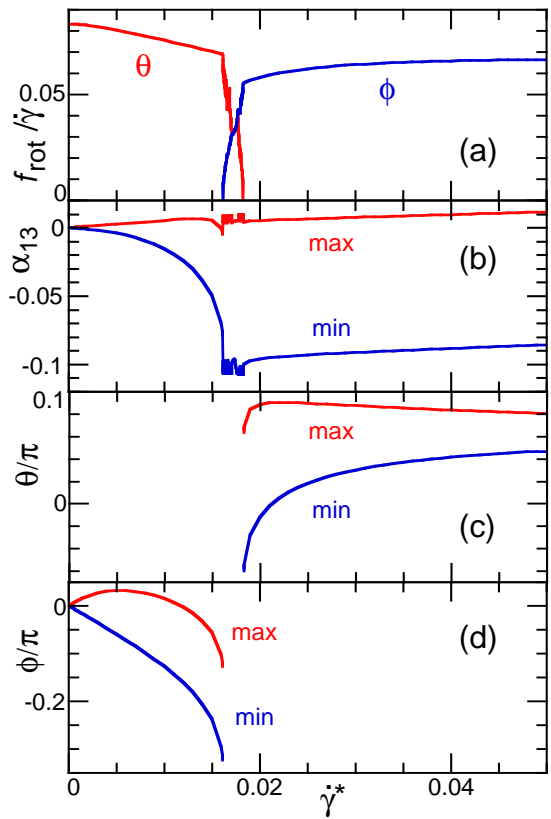}
\caption{ \label{fig:rq_mx}
(Color online)
Shear rate $\dot\gamma^*$ dependence 
of the deformable RBCs at $\eta_{\rm {in}}^*=0.5$.
(a) Rotation frequency $f_{\rm {rot}}$ of 
$\theta$ and $\phi$.
The maximum and minimum of (b) the shape parameter $\alpha_{13}$
, (c) $\theta$, and (d) $\phi$ are shown.
}
\end{figure}

\begin{figure}
\includegraphics{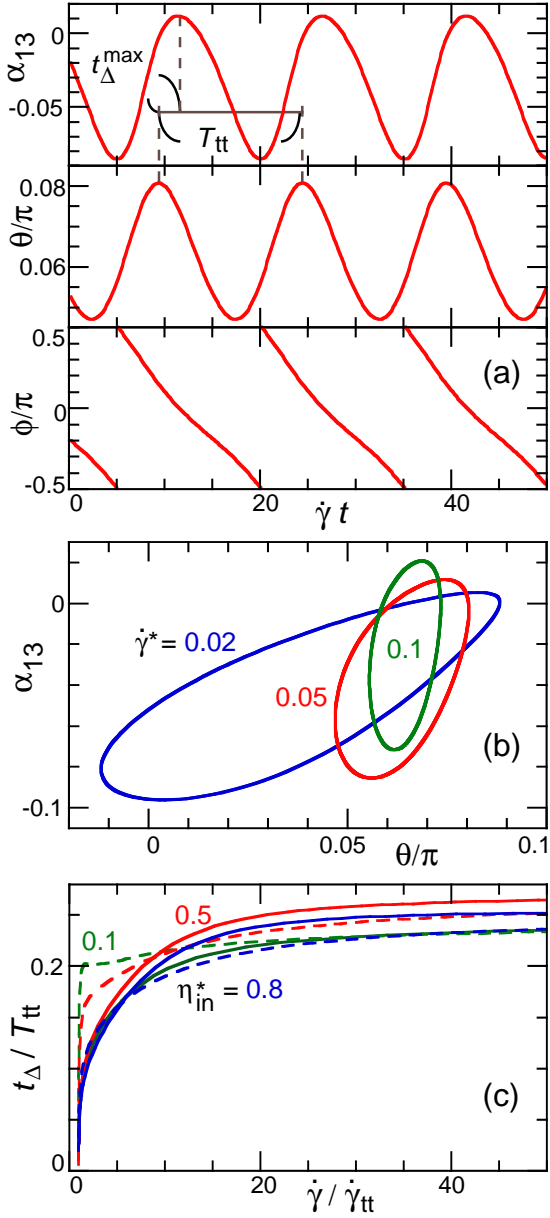}
\caption{ \label{fig:det_a11}
(Color online)
Dynamics of the deformable RBCs at $\eta_{\rm {in}}^*=0.5$.
(a) Time evolution of $\alpha_{13}$, $\theta$, and $\phi$ at $\dot\gamma^*=0.05$.
(b) Trajectories in the phase space ($\alpha_{13}$, $\theta$) at $\dot\gamma^*=0.02$, $0.05$, and $0.1$.
(c) Shear rate $\dot\gamma^*$ dependence of the phase difference between $\alpha_{13}$ and $\theta$
oscillations at $\eta_{\rm {in}}^*=0.1$.
Solid and dashed lines represent $t_{\Delta}^{\rm {max}}/T_{\rm {tt}}$
and $t_{\Delta}^{\rm {min}}/T_{\rm {tt}}$, respectively,
where $T_{\rm {tt}}$ is the TT rotation period, and
$t_{\Delta}^{\rm {max}}$ ($t_{\Delta}^{\rm {min}}$)
are the time difference between the maximum (minimum) of $\alpha_{13}$ and $\theta$;
see the explanation in (a).
}
\end{figure}

The free energy $F(\alpha_{13},\phi)$ is estimated by
the simulation of a model RBC with $\phi=0,\pi/2$ elongated by mechanical forces,
where the RBC membrane is modeled as a triangular network \cite{nogu09}.
In the simulation, $578$ vertices 
are connected by a bond potential $U_{\rm {bond}}=(k_1/2)(r-r_0)^2\{1+ (k_2/2)(r/r_0-1)^2\}$
with $\mu=(\sqrt{3}/4)k_1=6\times 10^{-6}N/m$, $\kappa=2 \times 10^{-19}$J, and  $k_2=1$.
The area and volume of the RBC are kept constant by harmonic potentials.
Our simulation reproduces the force-length curves of the optical-tweezers experiment \cite{mill04}
and other simulations \cite{dao06,dupi07,vazi08};
see Fig. 5 in Ref.~\cite{nogu09}.
The effective force  $\partial F(\alpha_{13},\phi)/\partial \phi$ is 
estimated from these force-length curves; see Fig.~\ref{fig:stretch}.
The model RBC at $\phi=\pi/2$ has $9\times 10^{-18}$J higher energy than at $\phi=0$
with $\alpha_{13}=-0.1$ in the absence of external forces.
This height of the energy barrier agrees with the value $E_0= 10^{-17}$J in Ref.~\cite{skot06},
which was estimated from Fischer's experiments \cite{fisc04}.
Abkarian {\it et~al.} estimated the height as
$E_0=\mu S (a_1/a_2- a_2/a_1)^2/2 \sim 3$ to $7 \times 10^{-10}$ [${\rm m}^2$] $\times \mu$
based on the velocity field of the KS theory [Eq. (\ref{eq:KS-vel})]~\cite{abka07}.
However, it gives much higher barrier $E_0 \sim 10^{-15}$J for $\mu=6\times 10^{-6}N/m$,
or smaller shear modules $\mu \sim 10^{-8}N/m$ for $E_0=10^{-17}$J.
Since the KS velocity field does not satisfy the local area conservation of membrane,
it may give more stress on the membrane than the area-conserving velocity field~\cite{seco82},
and the barrier height may be overestimated.

In this section, we employ the free-energy potential
  $F_0(\alpha_{13},\phi)=F_1(\alpha_{13}) + F_2(\alpha_{13})\sin^2(\phi)$.
The dependence on the potential function is discussed in Sec.~\ref{sec:pot}.
Instead of an interpolation~\cite{nogu04,nogu05,nogu07b},
we used fit functions to obtain smooth functions for the numerical calculations: 
The normalized potentials
 $F^*_1(\alpha_{13})= 5\alpha_{13}^2+ (40/3)\alpha_{13}^3+
  (230/4)\alpha_{13}^4$, $F^*_2(\alpha_{13})= 0.2+0.8\alpha_{13}$;
the shape
  parameter $\alpha_{12}=0.56+0.35\alpha_{13} -0.23\alpha_{13}^2
  +0.034\alpha_{13}^3$; the coefficients $f_2= 0.6018+0.064\alpha_{13}
  -0.19\alpha_{13}^2-0.42\alpha_{13}^4$, $f_3= 0.734+0.54\alpha_{13} +
  0.91\alpha_{13}^2+3.2\alpha_{13}^4$; and $\alpha_{13}^{\rm {max}}=0.7$.
Eqs.~(\ref{eq:qks}), (\ref{eq:phiks}), and (\ref{eq:al})
 are numerically integrated
using the fourth-order Runge-Kutta method.

In this model,
the viscosity ratio of the membrane and inner fluid and the reduced volume
are fixed at $\eta_{\rm {mb}}^*/\eta_{\rm {in}}^*=3.1$ and $V^*=V/(4\pi R_0^3/3)=0.64$.
Experimentally, the viscosity $\eta_0$ of outside fluid is typically varied
and $\eta_{\rm {mb}}$ and $\eta_{\rm {in}}$ are fixed at physiological values.
Thus, $\eta_{\rm {in}}^*=1$ corresponds to $\eta_0= 0.01$Pa$\cdot$s.
The reduced shear rate $\dot\gamma^*=1$ corresponds to $\dot\gamma=90 {\rm s}^{-1}$ or $180 {\rm s}^{-1}$,
at $\eta_{\rm {in}}^*=0.5$ or $1$, respectively.

\begin{figure}
\includegraphics{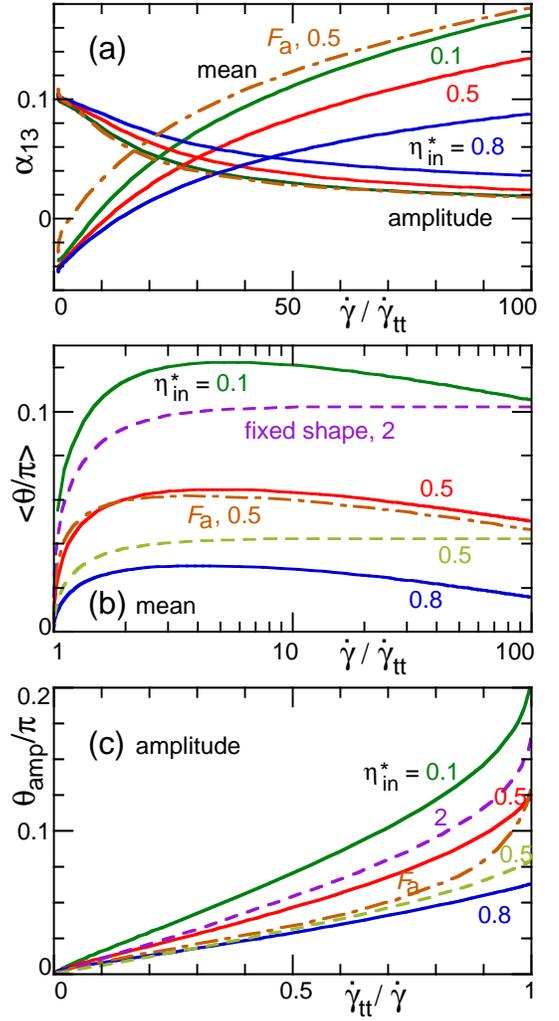}
\caption{ \label{fig:rq_tt}
(Color online)
Time averages and peak-to-peak amplitudes of the (a) $\alpha_{13}$ and (b), (c) $\theta$ oscillations
during TT motion.
Solid lines represent the deformable RBCs with the potential $F_0$ 
at $\eta_{\rm {in}}^*=0.1$, $0.5$, and $0.8$.
Dashed-dotted lines represent the deformable RBCs with the potential $F_{\rm a}$ 
at  $\eta_{\rm {in}}^*=0.5$.
Dashed lines in (b) and (c) represent the RBCs with the fixed shape at $\eta_{\rm {eff}}^*=0.5$ and $2$.
The shear rate  $\dot\gamma$ is normalized by the minimum shear rate $\dot\gamma_{\rm {tt}}$ of the TT phase.
}
\end{figure}

\begin{figure}
\includegraphics{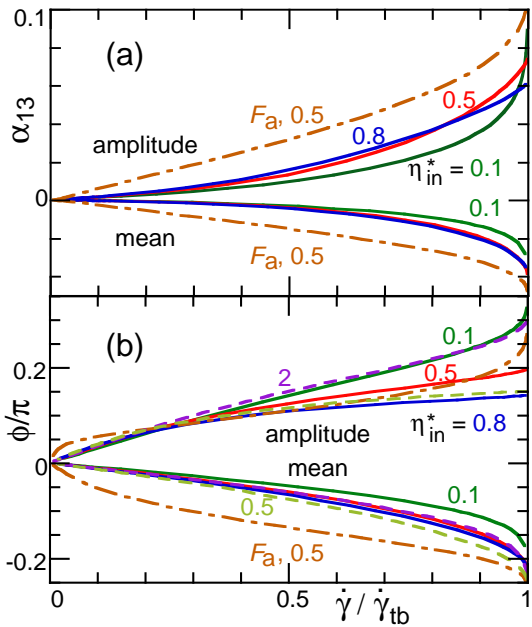}
\caption{ \label{fig:rq_tb}
(Color online)
Time averages and peak-to-peak amplitudes of the (a) $\alpha_{13}$ and (b) $\phi$ oscillations
during TB motion.
Solid lines represent the deformable RBCs with the potential $F_0$ 
at $\eta_{\rm {in}}^*=0.1$, $0.5$, and $0.8$.
Dashed-dotted lines represent the deformable RBCs with the potential $F_{\rm a}$ 
at  $\eta_{\rm {in}}^*=0.5$.
Dashed lines in (b) represent the RBCs with the fixed shape at $\eta_{\rm {eff}}^*=0.5$ and $2$.
The shear rate  $\dot\gamma$ is normalized by the maximum shear rate $\dot\gamma_{\rm {tb}}$ of the TB phase.
}
\end{figure}

\subsection{Results}

The phase diagram and rotation frequencies of the deformable RBC in simple flow are shown in Fig.~\ref{fig:phase_std}.
The shape deformation does not qualitatively change the phase diagram.
The tumbling phase appears in the bottom of Fig.~\ref{fig:phase_std}(a), 
since $\dot\gamma^*$ is used for the vertical axis instead of $U_{\rm e}\propto 1/\dot\gamma^*$.

In the TT phase, the shape parameter $\alpha_{13}$ and $\theta$ oscillate with the frequency $f_{\rm {rot}}^{\phi}$; 
see  Figs.~\ref{fig:rq_mx}(b) and \ref{fig:det_a11}(a).
As $\dot\gamma^*$ increases,
the time-average $\langle \alpha_{13} \rangle$ increases, and
the SW oscillation amplitudes of $\alpha_{13}$ and  $\theta$ decrease;
 see  Fig.~\ref{fig:rq_tt}(a).
The peak-to-peak amplitude $\theta_{\rm {amp}}$ is inversely proportional to $\dot\gamma^*$
for both the deformable and fixed-shape RBCs.
For the fixed-shape RBCs,
the mean angle $\langle \theta \rangle$ increases with increasing $\dot\gamma^*$
and reaches the angle of the KS theory at $U_{\rm e}=0$ ($\dot\gamma^* \to\infty$). 
For the deformable RBCs, $\langle \theta \rangle$ has a maximum and then decreases
because of the elongation of the RBCs;
 see  Figs.~\ref{fig:rq_tt}(b).

In the TT phase, 
$\alpha_{13}$ and $\theta$ oscillate with a fixed phase difference; see Fig.~\ref{fig:det_a11}.
The phase difference is calculated from the time difference $t_{\Delta}^{\rm {max}}$ between the maximum values of  $\alpha_{13}$ and  $\theta$
and from $t_{\Delta}^{\rm {min}}$ between the minimum values.
The difference between $t_{\Delta}^{\rm {max}}/T_{\rm {tt}}$ and $t_{\Delta}^{\rm {min}}/T_{\rm {tt}}$ 
represents the asymmetry of the oscillation functions.
In a sinusoidal function,  $t_{\Delta}^{\rm {max}}=t_{\Delta}^{\rm {min}}$.
It is found that $\alpha_{13}$ and $\theta$ show in-phase oscillation ($t_{\Delta}/T_{\rm {tt}}\simeq 0$) 
at small $\dot\gamma^*$,
and the phase difference approaches $\pi/4$ with increasing  $\dot\gamma^*$.
Walter {\it et al.} experimentally observed a phase difference of $\pi/4$ for synthetic capsules; 
see Fig. 7 in Ref.~\cite{walt01}.
Our results agree with their experiments.

In the TB phase,
$\langle \phi \rangle$ decreases and $\phi_{\rm {amp}}$ increases with increasing $\dot\gamma^*$; see Fig.~\ref{fig:rq_tb}.
When the energy barrier at $\phi/\pi=-0.5$ is overcome at $\dot\gamma_{\rm {tb}}^*$,
$\phi$ begins to rotate.
The average  $\langle \alpha_{13} \rangle$ decreases, since $F_0(\alpha_{13},\phi)$ has a minimum at $\alpha_{13}<0$ for $\phi\ne 0$.
In TB,
there is no significant difference between the deformable and fixed-shape RBCs.

\section{Dependence on potential function}~\label{sec:pot}

\begin{figure}
\includegraphics{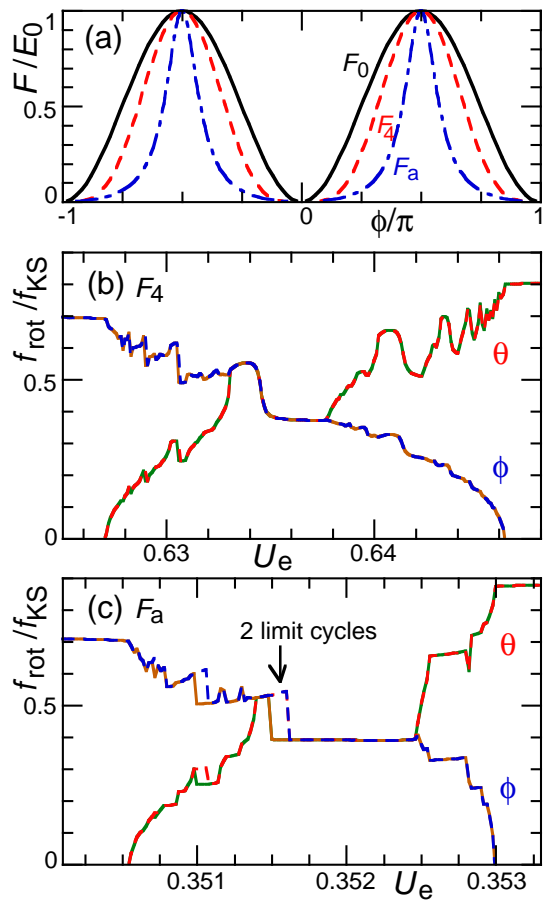}
\caption{ \label{fig:frot_s4}
(Color online)
(a) Three potentials $F_0=E_0\sin^2(\phi)$, $F_4=E_0\sin^4(\phi)$, 
and $F_{\rm a}=E_0\sin^2(\varphi)$.
The rotation frequencies $f_{\rm {rot}}$ of 
 $\theta$ and $\phi$ for fixed-shape capsules with (b) $F_4$ and (c) $F_{\rm a}$
at $\eta_{\rm {eff}}^*=1$.
Dashed or solid lines in (b) represent data
obtained with increasing or decreasing $\dot\gamma^*$, respectively.
The other parameters are the same as in Fig.~\ref{fig:skot_frot}.
}
\end{figure}

We compared the dynamics of RBCs with the fixed and deformable shapes in the previous section.
The fixation of the RBC shape can be interpreted as the bending rigidity $\kappa \to \infty$,
or $F(\alpha_{13},\phi)= k{\alpha_{13}}^2 + F_{\phi}(\phi)$ with $k\to \infty$.
This difference of the potential functions in $\alpha_{13}$ does not change the dynamics greatly.
In this section, we investigate the dependence on the potential functions in $\phi$.
For fixed-shape capsules, we compare the dynamics with three potential functions:
$F_0=E_0 \sin^2(\phi)$, $F_4=E_0\sin^4(\phi)$, and $F_{\rm a}=E_0 \sin^2(\varphi)$; see Fig.~\ref{fig:frot_s4}(a).
The angle $\varphi$ is the rotational angle $\varphi=\arctan(x_i/y_i)$,
where ${\bold r}_i=(x_i,y_i,z_i)$ is the position of a tracer on the membrane.
The phase angle $\phi$ is defined as $\phi=\arctan(x_i L_2/y_i L_1)$.
The angle $\varphi$ can be defined without assuming an ellipsoidal shape.
The potential $F_{\rm a}$ has the sharpest peak.

As the peak of potentials ($F_0$, $F_4$, $F_{\rm a}$) sharpens,
the transition shear rates $\dot\gamma^*_{\rm {tb}}$ and  $\dot\gamma^*_{\rm {tt}}$ increase
($U_{\rm e}$ decreases), since
the maximum forces  $\partial F/\partial \phi$ increases at constant $E_0$.
Two limit cycles of the synchronized rotations can coexist for $F_4$ and $F_{\rm a}$,
while no coexistence is observed for $F_0$.
The capsules approach different limit cycles with increasing and decreasing $\dot\gamma^*$.
For  the synchronized rotations with $f_{\rm {rot}}^{\theta}:f_{\rm {rot}}^{\phi}=1:1$,
 a steeper but continuous change appears in the $f_{\rm {rot}}$ curve for $F_4$ than for $F_0$,
and a discrete change with hysteresis appears for $F_{\rm a}$;
compare Figs.~\ref{fig:skot_frot}(a) and Figs.~\ref{fig:frot_s4}(b), (c). 

\begin{figure}
\includegraphics{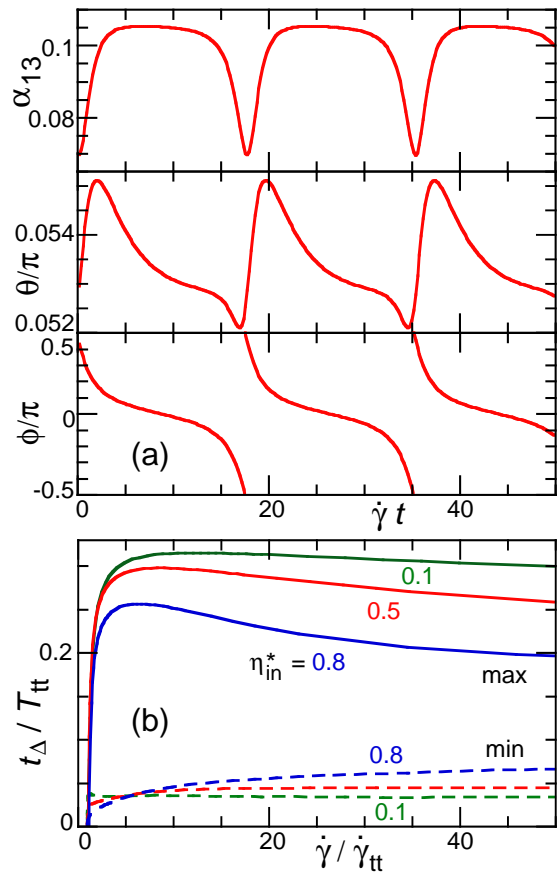}
\caption{ \label{fig:phi_det}
(Color online)
Dynamics of the deformable RBCs with the potential $F_{\rm a}$.
(a) Time evolution of $\alpha_{13}$, $\theta$, and $\phi$ at $\dot\gamma^*=1$ and  $\eta_{\rm {in}}^*=0.5$.
(b) Solid and dashed lines represent $t_{\Delta}^{\rm {max}}/T_{\rm {tt}}$
and $t_{\Delta}^{\rm {min}}/T_{\rm {tt}}$, respectively.
}
\end{figure}

The coexistence of two synchronized rotations occurs for the deformable RBCs with
the free energy $F_{\rm a}(\alpha_{13},\phi)=F_1(\alpha_{13}) + F_2(\alpha_{13})\sin^2(\varphi)$.
The shape parameter $\alpha_{13}$ and $\theta$ steeply change at $\phi \simeq \pm 0.5\pi$ during TT motion,
as shown in Fig.~\ref{fig:phi_det}.
The oscillations of $\alpha_{13}$ and $\theta$ do not approach sinusoidal curves, unlike the case for $F_0$;
compare Figs.~\ref{fig:det_a11}(c) and \ref{fig:phi_det}(b).
In the TT and TB phases,
the averages and amplitudes of the angles $\theta$ and $\phi$
do not show significant differences between the potential shapes in $\phi$ ($F_0$, $F_4$, and $F_{\rm a}$),
and in $\alpha_{13}$ (fixed or deformable shape); see Figs.~\ref{fig:rq_tt} and \ref{fig:rq_tb}.

\section{Discussion}~\label{sec:dis}

Let us compare our results with the experimental results~\cite{abka07}.
The TT-TB transition is observed at $\dot\gamma^*=0.01 \sim 0.04$ for $\eta_{\rm {in}}^*=0.45$
in Ref.~\cite{abka07}.
It is in good agreement with our results,
 $\dot\gamma^*_{\rm {tb}}=0.0158455$ and $\dot\gamma^*_{\rm {tt}}=0.0173455$ with $F_0$
and $\dot\gamma^*_{\rm {tb}}=0.0272$ and $\dot\gamma^*_{\rm {tt}}=0.028$ with $F_{\rm a}$.
Abkarian {\it et al.} also reported the dependences in the TT phase,
$f_{\rm {rot}} \propto \dot\gamma$ ($T_{\rm {tt}}\propto 1/\dot\gamma$) and the SW amplitude 
$\theta_{\rm {amp}}\propto 1/\dot\gamma$.
We also obtained these dependences for all types of potentials; 
see  Figs.~\ref{fig:rq_mx}(a) and \ref{fig:rq_tt}(c).
Most of the dynamics in our calculations are qualitatively independent of the potential shapes,
and the quantitative differences are smaller than the distribution widths of the experimental data.
In the experiments, the transition shear rate and the other quantities have wide distributions
because of the polydispersity of RBCs.
RBCs become smaller and more viscous with age~\cite{tran84,nash83}.
The viscoelasticity of the RBC membrane is changed
by some diseases such as diabetes mellitus \cite{tsuk01}.
Further experimental and simulation studies are needed
to tune up the RBC potential in our theoretical model.

In experiments, it is difficult to distinguish
the intermittent rotations from transient rotations~\cite{abka07}.
The intermittency has not yet been obtained by numerical simulations~\cite{kess08,sui08}.
The synchronized rotations, in particular with $f_{\rm {rot}}^{\theta}:f_{\rm {rot}}^{\phi}=1:1$,
would be much easier to observe in experiments and simulations.
Kessler {\it et al.}~\cite{kess08} argued that the intermittent phase might be an artifact of the theory,
since they did not obtain it in their simulations.
In our study, however, 
the modifications of the theory do not qualitatively change the phase diagram.
The Devil's staircase is the general dynamics on a torus (two-dimensional plane with periodic boundary condition).
We believe that the 
intermittent and synchronized rotations occur at higher $\dot\gamma$ than their simulated values.

Interestingly, for TT motion slightly above $\dot\gamma_{\rm {tt}}^*$,
 the angle $\theta$ oscillates crossing $\theta=0$;
 see  Figs.~\ref{fig:rq_mx}(c).
Fluid vesicles also show SW oscillation of the vesicle shape and $\theta$ around $\theta \simeq 0$ 
between the TT and TB phases by a different mechanism,
where the shrinkage of the vesicles at $\theta<0$ induces a change of the dynamic mode from TB ($B<1$) to TT ($B>1$)
in the generalized KS theory~\cite{nogu07b}.
Previously, we distinguished that
in the SW oscillation of the elastic capsules, $\theta$ is always positive and the shape deformation is negligibly small,
while in SW of fluid vesicles, $\theta$ changes its sign and the shape shows large deformation.
However, we know now that the condition for $\theta$ is not always true.
The clear difference is the dependence on $\eta_{\rm {in}}^*$ or  $\eta_{\rm {mb}}^*$.
SW induced by the shape deformation appears only in a narrow range of the viscosity,
whereas SW induced by the membrane shear elasticity appears at a wide range of the viscosities with no lower viscosity limit.
In the future, it will be interesting to investigate the coupling of different 
oscillation mechanisms in elastic capsules.

\section{Summary}~\label{sec:sum}

In summary, we described the dynamics of RBCs in simple shear flow using a simple theory.
The phase diagram of RBCs is divided  into three regions:
tank-treading, tumbling, and intermediate regions.
In the intermediate regions,
RBCs exhibit intermittent or synchronized rotations of  the inclination angle $\theta$ and  phase angle $\phi$. 
Synchronized rotations, in particular with $f_{\rm {rot}}^{\theta}:f_{\rm {rot}}^{\phi}=1:1$,
would be much easier to experimentally observe than intermittent rotations.
In the TT (TB) phase, the shape and $\theta$ ($\phi$) oscillate with the frequency of $\phi$ ($\theta$) rotation.
The coexistence of two  synchronized rotations can appear when the potential function of $\phi$ has a sharp peak.
The other dynamic properties  are not sensitive to the function shape of the free-energy potential.
We focused on the dynamics of RBCs in this paper, but 
 the resulting dynamics would be generally applicable to other elastic capsules.

\begin{acknowledgments}
We  would like to thank 
G. Gompper (J{\"u}lich) for the helpful discussion.
This study is partially supported by a Grant-in-Aid for Scientific Research on Priority Area ``Soft Matter Physics'' from
the Ministry of Education, Culture, Sports, Science, and Technology of Japan.
\end{acknowledgments}

\bibliographystyle{apsrev}
%\bibliography{tri}

\end{document}